\documentclass[prl,twocolumn,noshowpacs,preprintnumbers,amsmath,amssymb,linenumbers]{revtex4}

\usepackage{mathrsfs}
\usepackage[dvipdfmx]{graphicx}
\usepackage{dcolumn}
\usepackage{bm}
\usepackage{color}
\usepackage{times}
\usepackage{ulem}

\newcommand{\tb}{\textbf}

\newcommand{\ra}{\rangle}
\newcommand{\la}{\langle}

\newcommand{\eps}{\epsilon}

\newcommand{\ST}{{\rm ST}}

\newcommand{\unbiased}{{\rm unbiased}}
\newcommand{\naive}{{\rm NLLS}}

\begin{document}
\title{
	Can we infer microscopic financial information from the long memory in market-order flow?:\\ 
	a quantitative test of the Lillo-Mike-Farmer model
}

\author{Yuki Sato}
\affiliation{
	Department of Physics, Graduate School of Science, Kyoto University, Kyoto 606-8502, Japan
}
\author{Kiyoshi Kanazawa}
\email{kiyoshi@scphys.kyoto-u.ac.jp}
\affiliation{
	Department of Physics, Graduate School of Science, Kyoto University, Kyoto 606-8502, Japan
}	
\date{\today}

\begin{abstract}
 In financial markets, the market order sign exhibits strong persistence, widely known as the long-range correlation (LRC) of order flow; specifically, the sign autocorrelation function (ACF) displays long memory with power-law exponent $\gamma$, such that $C(\tau) \propto \tau^{-\gamma}$ for large time-lag $\tau$. One of the most promising microscopic hypotheses is the order-splitting behaviour at the level of individual traders. Indeed, Lillo, Mike, and Farmer (LMF) introduced in 2005 a simple microscopic model of order-splitting behaviour, which predicts that the macroscopic sign correlation is quantitatively associated with the microscopic distribution of metaorders. While this hypothesis has been a central issue of debate in econophysics, its direct quantitative validation has been missing because it requires large microscopic datasets with high resolution to observe the order-splitting behaviour of all individual traders. Here we present the first quantitative validation of this LMF prediction by analysing a large microscopic dataset in the Tokyo Stock Exchange market for more than nine years. On classifying all traders as either order-splitting traders or random traders as a statistical clustering, we directly measured the metaorder-length distributions $P(L)\propto L^{-\alpha-1}$ as the microscopic parameter of the LMF model and examined the theoretical prediction on the macroscopic order correlation $\gamma \approx \alpha - 1$. We discover that the LMF prediction agrees with the actual data even at the quantitative level. We also discuss the estimation of the total number of the order-splitting traders from the ACF prefactor, showing that microscopic financial information can be inferred from the LRC in the ACF. Our work provides the first solid support of the microscopic model and solves directly a long-standing problem in the field of econophysics and market microstructure.
\end{abstract}

\maketitle

\paragraph{Introduction.}
	Can a statistical-physics approach help in understanding macroscopic phenomena in financial markets from their microscopic dynamics~\cite{EconPhys_Stanley,EconPhys_Slanina}? In posing this challenging thought, physicists have greatly benefitted from recent high-frequency data for econophysics modelling of market microstructure~\cite{BouchaudText,PhysRep2022}, even at the level of individual traders~\cite{KanazawaPRL,KanazawaPRE}. In this Letter, we provide the first quantitative evidence of a historic econophysics theory regarding the long-range correlation (LRC) in the market order flow~\cite{LMF,Bouchaud2003,Lillo2004}.

	Let us briefly review the trading rules in recent financial markets, where traders have two options. The first option is the limit order, by which traders provide the market liquidity and show the potential prices at which they are willing to transact. The second option is the market order, by which traders immediately consume the liquidity and transact at the best prices (i.e., the highest bid or the lowest ask prices). This Letter tests an econophysics microscopic model for the market-order flow, particularly on their statistical persistence.

\begin{figure*}
		\centering
		\includegraphics[width=140mm]{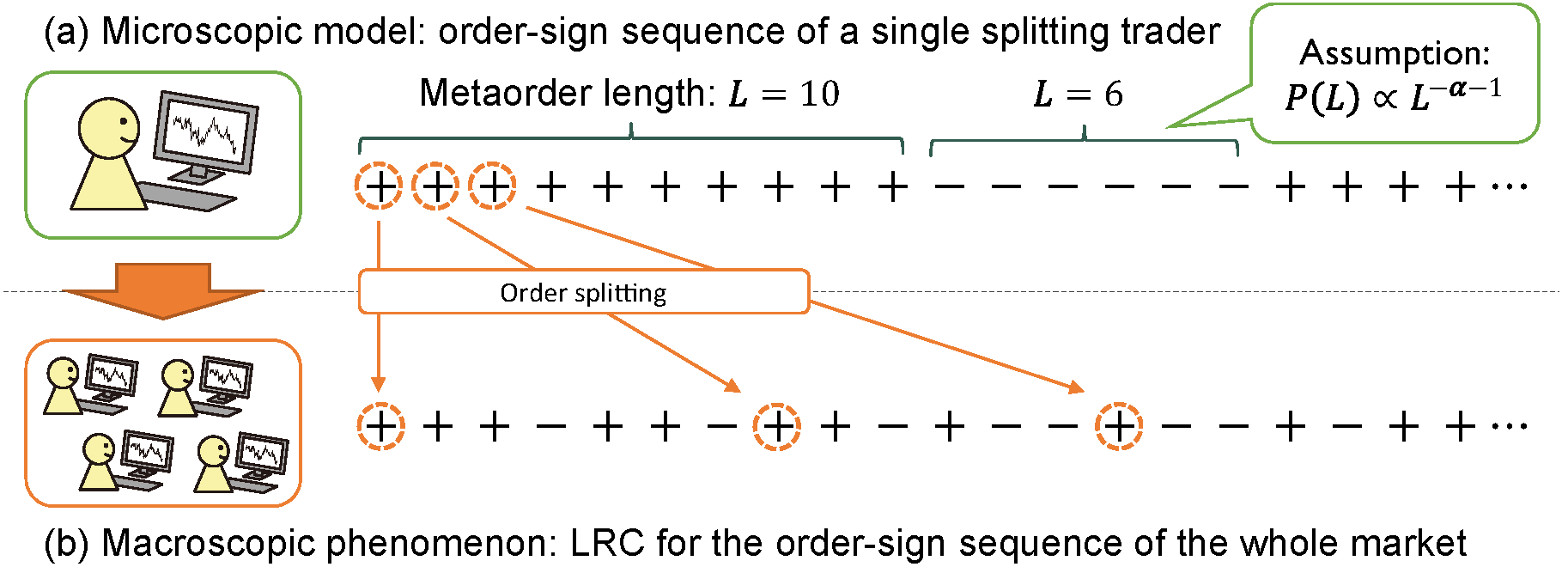}
		\caption{
			Schematic of the LRC of the market-order flow and the order-splitting hypothesis (particularly, the LMF model); as a shorthand notation for $+1$ ($-1$), ``$+$'' (``$-$'') signifies a buy (sell). 
			(a)~As a microscopic model, we assume the presence of STs. Also, STs successively submit the child orders with the same sign for $L$ times, where $L$ is called the metaorder length and obeys power law statistics, $P(L)\propto L^{-\alpha-1}$. 
			(b)~Consequently, the LRC appears as a macroscopic phenomenon. The LMF theory predicts a quantitative relation $\gamma=\alpha-1$, which we empirically establish in this Letter through data analysis.
		}
		\label{fig:LMF}
	\end{figure*}

	The strong persistence of the market-order flow underscores an established empirical law in financial markets~\cite{BouchaudText,Bouchaud2003,Lillo2004}: i.e., once a buy (sell) market order is observed, a buy (sell) market order is likely to be observed (see Fig.~\ref{fig:LMF}). This predictability regarding market orders is mathematically characterised by a power-law decay for the order-sign autocorrelation function (ACF):
	\begin{equation}
		C(\tau) := \la \eps(t)\eps(t+\tau) \ra \approx c_0 \tau^{-\gamma}, \>\>\> 0 < \gamma < 1,
		\label{eq:LRC_ACF}
	\end{equation}
	for large time-lag $\tau \gg 1$. Here $\eps(t)$ is the market order sign at time $t$ defined by $\eps(t)=+1$ ($\eps(t)=-1$) for the buy (sell) market order, $\la\dots \ra$ represents the ensemble average, $c_0$ is the prefactor, and $\gamma$ is the power-law exponent for the LRC. Because it is ubiquitously observed across broad markets, the LRC is believed essential to a market microstructure.
	
	Then, what is the microscopic origin of the LRC as a macroscopic phenomenon? One promising response is the order-splitting hypothesis for individual traders' behaviours~\cite{LMF} (Fig.~\ref{fig:LMF}(a)). This hypothesis claims the LRC appears because some traders split large metaorders into a long series of small child orders. Because all the child orders share for a while the same sign, there is weak predictability of the future order sign, which is ultimately reflected in the power-law decay of the ACF as summarised in Eq.~\eqref{eq:LRC_ACF} (Fig.~\ref{fig:LMF}(b)). Furthermore, Lillo, Mike, and Farmer (LMF) proposed a simple microscopic theory based on the order-splitting hypothesis. They assumed (i)~the presence of splitting traders (STs), and (ii)~the power-law probability density function (PDF) for the metaorder length $L$ such that $P(L)\propto L^{-\alpha-1}$ with microscopic exponent $\alpha>1$. By assuming random order submissions, the ACF macroscopically exhibits a power-law decay~\eqref{eq:LRC_ACF}. Specifically, they showed
	\begin{equation}
		\gamma = \alpha - 1,
		\label{eq:quantitative_pred}
	\end{equation}
	which in this Letter we refer to as the {\it quantitative LMF prediction}. The prediction~\eqref{eq:quantitative_pred} is beautiful and quantitatively powerful because it connects the macroscopic and microscopic parameters in alignment with the central spirit of statistical physics. 

	While the plausibility of this scenario was confirmed qualitatively in \cite{Toth2015} (i.e., a decomposition of the ACF into an order-splitting component and the remainder), the detailed verification of the quantitative prediction~\eqref{eq:quantitative_pred} has been missing for 18 years. The original LMF paper~\cite{LMF} reported an initial attempt to test their prediction. However, they only confirmed a minimum consistency of their theory (i.e., the theoretical line passes through the centre of the mass in the scatterplot; see Fig. 5 and Sec.~III B in \cite{PRR_companion} for a brief review) when lacking suitably large datasets. 

	In this Letter, together with companion paper~\cite{PRR_companion}, we solve this long-standing econophysics problem precisely by analysing a large comprehensive microscopic dataset of the Tokyo Stock Exchange (TSE). We accessed a special microscopic dataset, including trading-account identifiers (IDs) on the TSE, enabling us to track effectively the behaviour of trading accounts. Using our microscopic dataset, we first applied a strategy clustering of individual traders to test assumption~(i). In regard to market orders, and after classifying all traders as STs or random traders (RTs), we confirmed the presence of STs in most of the TSE markets. We next studied the empirical metaorder-length PDF $P(L)$ to test assumption~(ii), which we validated from our dataset. With the measured microscopic parameter $\alpha$, we generated a scatterplot between $\alpha$ and $\gamma$ to test the quantitative LMF prediction~\eqref{eq:quantitative_pred}. Finally, we found the prediction~\eqref{eq:quantitative_pred} agreed with our dataset, providing quantitatively the first solid support for the LMF model as the minimal microscopic description of the order-splitting behaviour. As the last discussion, we estimate the total number of the STs from the observed prefactor $c_0$. Our findings imply that the long memory in the market-order ACF is useful in inferring microscopic financial information.

\paragraph{Data description.}
	Let us briefly describe our dataset provided by the Japan Exchange (JPX) Group, the platform manager of the TSE. The TSE being the biggest stock market in Japan, our dataset covers all the order flows in the TSE (market orders, limit orders, and cancellations), enabling us to track their complete life cycle for all the stocks for nine years (from the 4th January 2012 to the 30th December 2020). Furthermore, this dataset includes virtual server IDs (VSIDs), a unit of trading accounts on the TSE. The VSID is not technically equivalent to the membership ID, because any trader may have several VSIDs. However, we can effectively define trader IDs to track individual trader behaviour with high resolution by appropriately aggregating VSIDs~\cite{TradingDesks1,TradingDesks2} (e.g., if a limit order is submitted by VSID 1 and is cancelled from VSID 2, both VSIDs are associated with the same trader); see also \cite{PRR_companion} for more technical details. 

	Our study focused on the sign sequences of market orders during double auctions from 09:00-11:30 and 12:30-15:00 Japan Standard Time. A yearly segmented order-sign sequence was extracted for each stock to obtain one market datapoint. We only used datapoints with more than 0.5 million transactions and removed transaction data from the opening and closing ten minutes of auctions to suppress intraday-seasonality effect. 

\paragraph{Assumptions of the LMF model.}
	As summarised in Fig.~\ref{fig:LMF}, there are two key assumptions in the LMF model: (i) the presence of STs who have large latent demand ({\it metaorders}) and split them into small {\it child orders}, which are assumed to share the same sign for $L$ successive times, and (ii) the {\it metaorder length} $L$ obeys a power law $P(L)\propto L^{-\alpha-1}$ with $\alpha>1$.

	In previous literature, there was no solid direct evidence of assumption~(i), although \cite{Toth2015} shows indirect but promising evidence based on the ACF decomposition. Also, the plausibility of assumption~(ii) was studied in \cite{LMF} by analysing the off-book data for the London stock exchange market as an ``imperfect proxy". However, with the absence of appropriate datasets at that time, the precise estimation of $\alpha$ became a technical problem for LMF verification. To verify assumptions~(i) and (ii) directly, it is necessary to identify STs by strategy clustering at the level of individual traders and then study their metaorder-length PDF to measure $\alpha$ precisely. 

	\paragraph{Presence of STs.}
	\begin{figure}
		\centering
		\includegraphics[width=85mm]{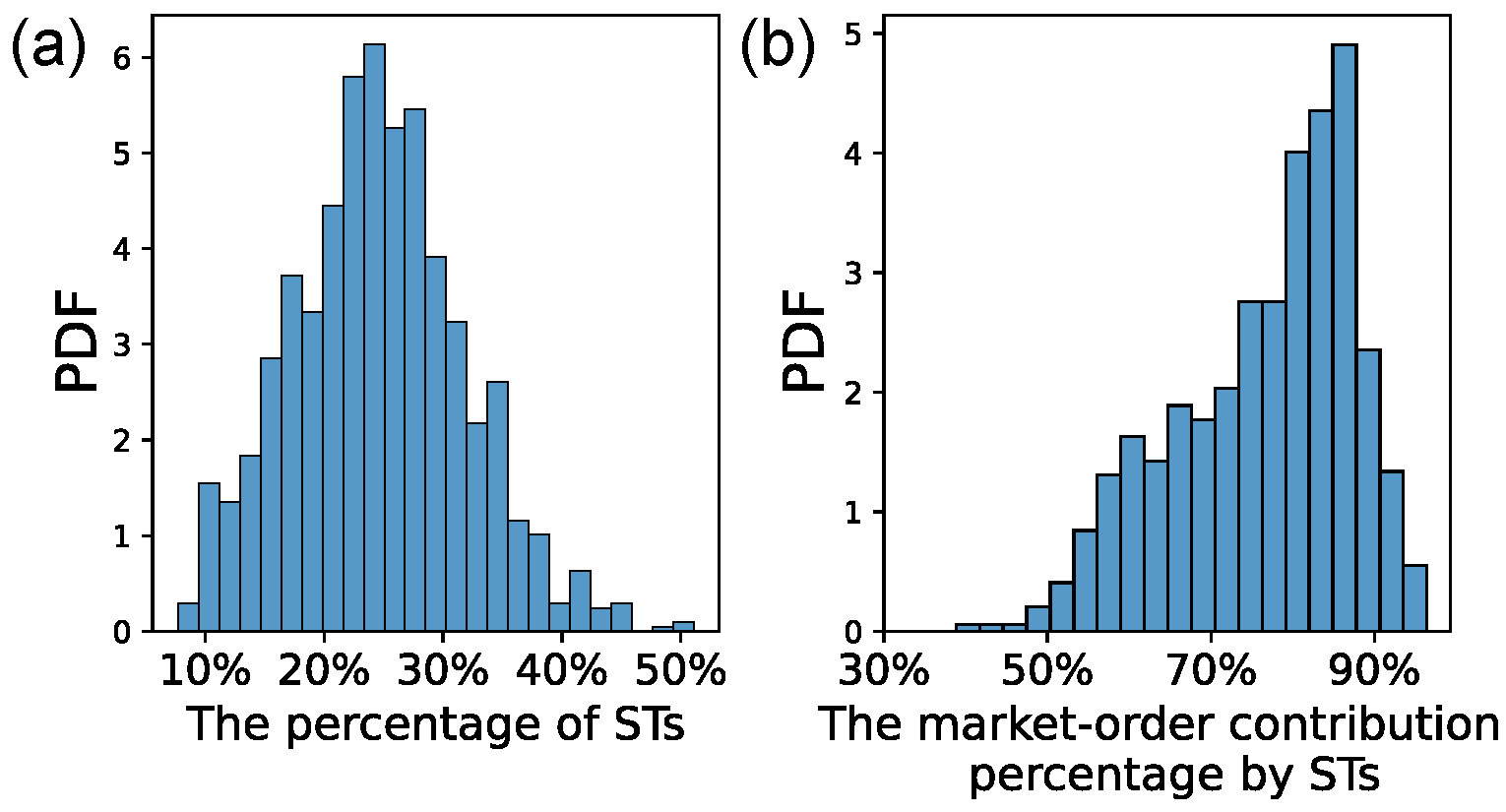}
		\caption{
			Presence of the STs by our strategy clustering. 
			(a)~Empirical PDF for the percentage of STs in each market, showing direct evidence of the presence of STs. Typically, 25\% of all traders were STs. 
			(b)~Empirical PDF for STs' contribution to market orders in each market. Typically, 80\% of all the market orders were issued by the STs, implying their overwhelming contribution to market orders.
		}
		\label{fig:ST_statistics}
	\end{figure}
	We proceeded with strategy clustering to identify STs. We studied the order-sign sequence for each ST (Fig.~\ref{fig:LMF}(a)) to construct the metaorder length $L$ by defining $L$ as a length of successively equal signs. Concerning exceptional handling, if there was more than one business day between two successive orders, we assume they belong to different metaorders~\cite{Donier2015} to avoid overestimating metaorder length. 

	For a given metaorder-length sequence, we apply the binomial test for strategy clustering; the null hypothesis is that the order-sign sequence is purely random (obeying a symmetric Bernoulli process) and, thus, the trader belongs to the RT set. The trader is regarded as an ST if the null hypothesis is rejected with a significance level $\theta:=0.01$.

	On the basis of this clustering scheme, we identified the ST set for each market datapoint. With summary statistics across all the markets during nine years, we evaluated the empirical PDF for the ST percentage in each market [Fig.~\ref{fig:ST_statistics}(a)], and the contribution to market orders from the ST set [Fig.~\ref{fig:ST_statistics}(b)]. We concluded that typically a quarter of all traders are STs, but they dominate the total market orders. Via this strategy clustering, we thus validated assumption~(i) directly.

\paragraph{Metaorder-length PDF.}	
	\begin{figure}
		\centering
		\includegraphics[width=85mm]{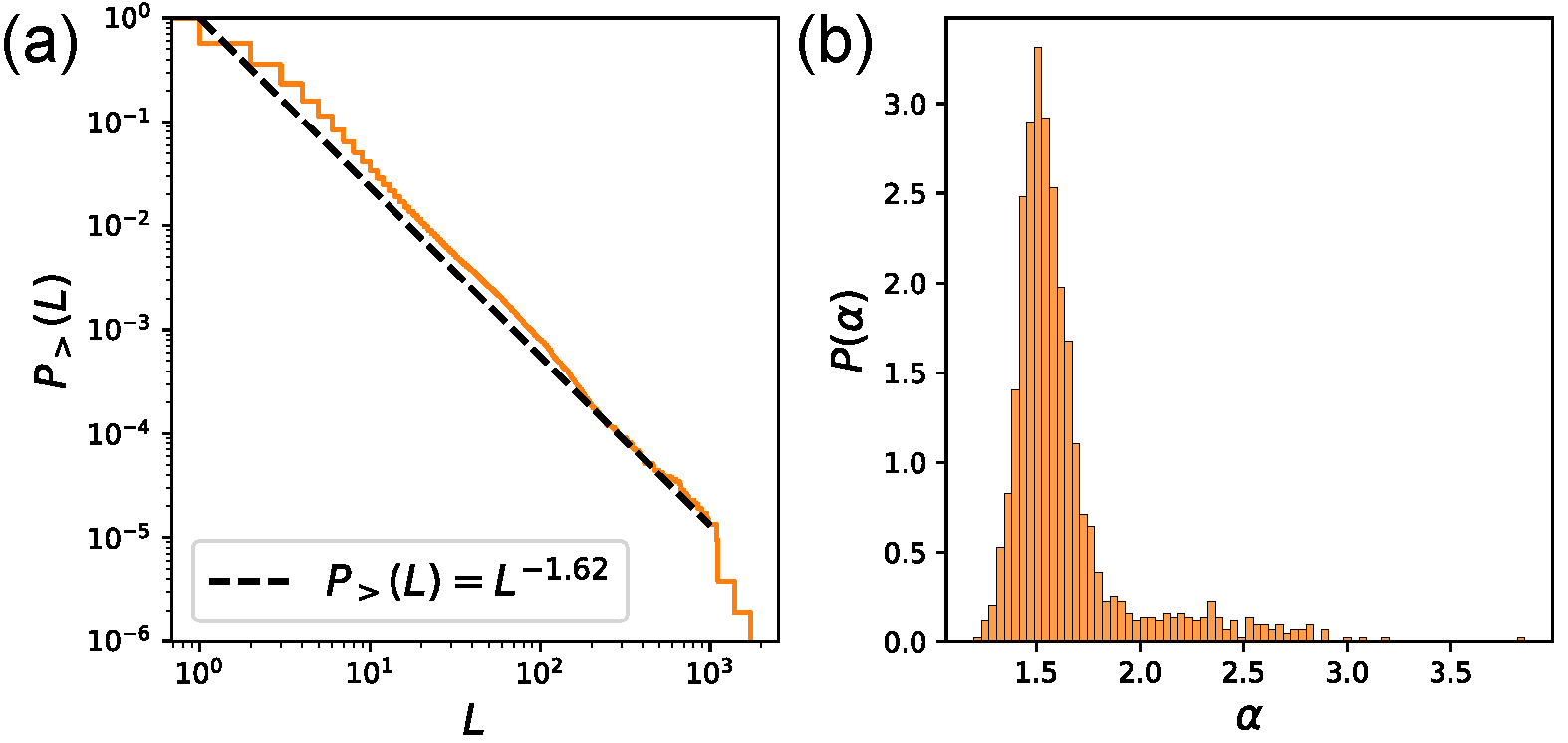}
		\caption{
			The aggregated CCDFs for STs with our strategy clustering. 
			(a)~Empirical CCDF aggregated regarding metaorder length $L$ among STs using data for the Toyota Motor Corporation in 2020 as a typical example. The CCDF obeys the power law $P_>(L)\propto L^{-\alpha}$ with $\alpha\approx 1.62$. Likewise, most empirical aggregated CCDF for STs obey similar power laws. 
			(b)~Empirical PDF of the power-law exponent $\alpha$ for all the markets. The power-law exponents were evaluated systematically using Clauset's algorithm~\cite{Clauset2009,Alstott2014} across all the markets. The exponent $\alpha$ typically satisfies $1 < \alpha < 2$, consistent with the standard assumption for the LMF model. 
		}
		\label{fig:MetaorderCCDFs}
	\end{figure} 
	Having identified the set of STs, we measured the aggregated empirical PDF for the metaorder length of all STs. Most of the aggregated complementary-cumulative distribution functions (CCDFs) for the metaorder length of STs obey a power law $P_>(L)\propto L^{-\alpha}$ with the CCDF defined by $P_>(L):=\int_{L}^\infty P(L')dL'$. As a typical example, we plotted the metaorder-length CCDF for the Toyota Motor Corporation (with ticker number 7203) in 2020 [Fig.~\ref{fig:MetaorderCCDFs}(a)]; it features a power-law asymptotic tail for large $L$. We then evaluated $\alpha$ using Clauset's algorithm~\cite{Clauset2009,Alstott2014} to plot the empirical PDF of $\alpha$ [Fig.~\ref{fig:MetaorderCCDFs}(b)] across all the stocks. Typically, the exponent $\alpha$ is distributed over values $1<\alpha<2$, in agreement with the standard assumption for the LMF model. We thus validated assumption~(ii) for our dataset. 

	\paragraph{Power-law exponent in the ACF.}
		Having measured the microscopic power-law exponent $\alpha$, we next measured the macroscopic power-law exponent $\gamma$ of the ACF, which we did by fitting directly the sample order-sign ACF as follows (see \cite{PRR_companion} for details of the method): We first calculated the sample ACF from $C_{\rm sample}(\tau):= \sum_{t=1}^{N_{\eps}-\tau}\eps(t)\eps(t+\tau)/(N_{\eps}-\tau)$ with time-lag $\tau$ and total number of market orders $N_{\eps}$. We fixed the fitting range $[\tau_{\rm th}^-, \tau_{\rm th}^+]$ automatically such that only the power-law decay is observed in the ACF for $[\tau_{\rm th}^-, \tau_{\rm th}^+]$. We applied logarithmic smoothing and a final fitting $C(\tau)=C_0\tau^{-\gamma_{\naive}}$ for the range $[\tau_{\rm th}^-, \tau_{\rm th}^+]$ employing the relative nonlinear least square (NLLS) estimation. 

		Although the NLLS estimator $\gamma_{\naive}$ gives numerical consistency for the LMF model, we noticed that the NLLS estimator $\gamma_{\naive}$ has a finite-sample-size bias. To remove this bias, we constructed heuristically an approximate unbiased estimator $\gamma_{\unbiased}$ based on the LMF model (see companion paper~\cite{PRR_companion} for details). For this Letter, we used this unbiased estimator $\gamma_{\unbiased}$ for the final scatterplot. 

		As a robustness check, we also measured the power-law exponent $\gamma$ via the power-spectral density (PSD) method (see \cite{PRR_companion}). The exponent measured by the ACF and PSD fittings are respectively denoted by $\gamma^{(a)}_{\unbiased}$ and $\gamma^{(s)}_{\unbiased}$. Both methods exhibit reasonable and consistent results, implying the statistical robustness of our results.

	\paragraph{Scatterplot.}
	\begin{figure*}
		\centering
		\includegraphics[width=175mm]{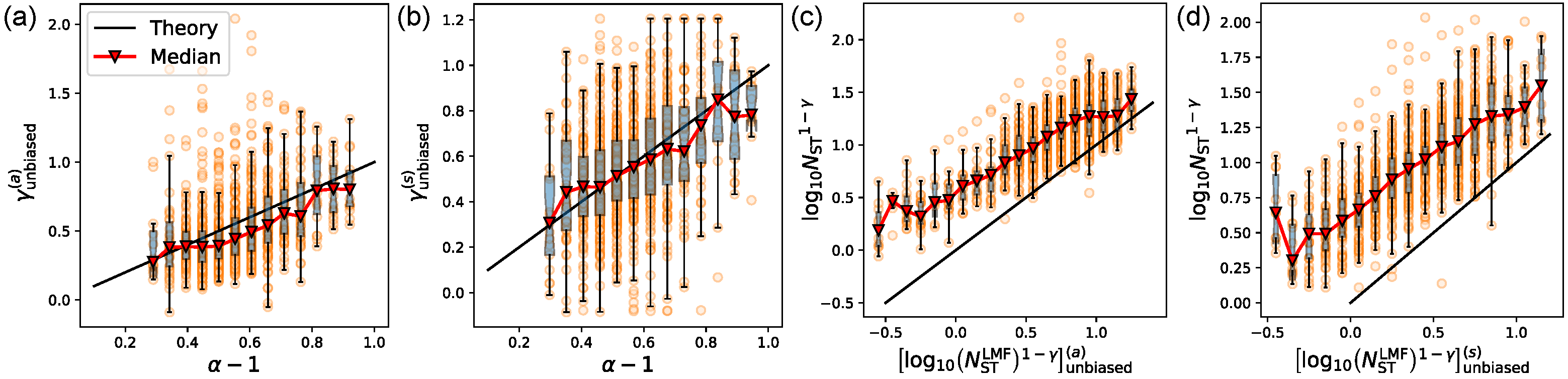}
		\caption{
			(a, b)~Scattered boxplots between $\alpha$ and $\gamma$ with the median, the first and third quartiles for (a) the ACF and (b) the PSD methods, exhibiting excellent agreement with the LMF prediction~\eqref{eq:quantitative_pred} (black line).
			$\gamma$ was evaluated using the approximate unbiased estimator $\gamma_{\unbiased}$, based on the NLLS estimator and the LMF model.	(c, d)~Scattered boxplots between the LMF estimator $N^{\rm LMF}_{\ST}$ and the actual total number of the STs $N_{\ST}$ for (c) the ACF and (d) the PSD methods for the datapoints with $\alpha<2$. The LMF estimator is highly correlated with the true value of $N_{\ST}$ as the classical theory predicts, but systematically underestimates $N_{\ST}$, such that $N^{\rm LMF}_{\ST}\lesssim N_{\ST}$. This observation is consistent with a generalised LMF theory~\cite{SatoJSP2023} with heterogeneous intensities $\{\lambda^{(i)}\}_{i}$.
		}
		\label{fig:scatterplot}
	\end{figure*}
	Having evaluated the microscopic and macroscopic power-law exponents $\alpha$ and $\gamma$ using our huge TSE dataset, we are ready to draw the scatterplot between $\alpha$ and $\gamma$ and test the LMF prediction~\eqref{eq:quantitative_pred}. As the main result, we provide the scattered boxplots (Figs.~\ref{fig:scatterplot}(a) and (b) for the ACF and PSD methods, respectively) between $\alpha$ and $\gamma_{\unbiased}$ with focus on the range $1<\alpha<2$ in accordance with the standard LMF assumption~\footnote{The scattered boxplot is composed of both boxplot and scatterplot. The datapoints are drawn on the centre of the bins along the $\alpha$-axis.}. These figures exhibit excellent agreement with the theoretical line~\eqref{eq:quantitative_pred}. From these figures, we conclude that with our microscopic dataset the LMF prediction~\eqref{eq:quantitative_pred} has quantitative validity.

\paragraph{Discussion on the prefactor.}
	While we extracted the microscopic information $\alpha$ from the ACF power-law exponent $\gamma$ via Eq.~\eqref{eq:quantitative_pred}, is it possible to extract other microscopic information from the prefactor $c_0$? The LMF theory predicts $c_0\simeq N_{\ST}^{\alpha-2}/\alpha$ with the total number of the STs $N_{\ST}$, implying that $N_{\ST}$ can be estimated by the LMF estimator 
	\begin{subequations}
		\label{eq:ACF_estimate_M_set}
		\begin{equation}
			N_{\ST}^{\rm LMF}\left(c_0, \gamma\right) := \frac{1}{\left[(\gamma+1)c_0\right]^{\frac{1}{1-\gamma}}},
			\label{eq:ACF_estimate_M}
		\end{equation}
		where $\gamma$ and $c_0$ can be observed from publicly data. 
		
		Note that original LMF work made an assumption of the homogeneiety of order-splitting intensities $\{\lambda^{(i)}\}_{i}$ among traders in \cite{LMF}, such that $\lambda^{(i)}=1/N_{\ST}$ for all $i$. While we noticed that this homogeneiety assumption is unrealistic, we tested this prediction in our dataset by drawing the scattered boxplots (Fig.~\ref{fig:scatterplot}(c) and (d) based on the ACF and PSD methods, respectively) between $\log_{10}N_{\ST}^{1-\gamma}$ and the LMF estimator $\log_{10}\left(N_{\ST}^{\rm LMF}\right)^{1-\gamma}$ with the finite-sample size bias removed (see Ref.~\cite{PRR_companion}). We find that the LMF estimator $N_{\ST}^{\rm LMF}$ is highly correlated with the true value $N_{\ST}$, implying that the ACF prefactor is a useful resource to infer $N_{\ST}$. At the same time, the LMF estimator $N_{\ST}^{\rm LMF}$ systematically underestimates the true value $N_{\ST}$, such that $N_{\ST}^{\rm LMF}\lesssim N_{\ST}$. 

		Interestingly, our finding is consistent with a generalised LMF model with the heterogeneous intensity distribution $\{\lambda^{(i)}\}_{i}$. Indeed, Ref.~\cite{SatoJSP2023} shows the ACF formula~\eqref{eq:ACF_estimate_M} is non-robust but sensitive to the heterogeneous intensity distribution, while the power-law-exponent formula~\eqref{eq:quantitative_pred} is robust. Furthermore, the LMF estimator $N_{\ST}^{\rm LMF}$ is shown to provide the lower bound of the true value of $N_{\ST}$, such that 
		\begin{equation}
			N_{\ST}^{\rm LMF}\lesssim N_{\ST},
			\label{eq:ACF_estimate_M_ineq}
		\end{equation}
	\end{subequations}
	showing the consistency with Fig.~\ref{fig:scatterplot}(c, d). Thus, we have successfully confirmed the qualitative validity of the LMF picture even for the estimation of $N_{\ST}$, while for better quantitative understanding it might require theoretical updates regarding the heterogeneity of trading strategies.

\paragraph{Conclusion.}
	Although the power-law memory character in the order-sign ACF has been a central issue in econophysics, and with the absence of an appropriate huge microscopic dataset, no quantitative evidence had been provided for the corresponding microscopic model (the LMF model). In this Letter, we have provided the first solid evidence for the LMF model at the quantitative level~\eqref{eq:quantitative_pred} at least for the TSE market and, thus, solved this long-lasting problem.

	Let us briefly discuss the implication of our findings. Our work shows that the microscopic parameters $\alpha$ and $N_{\ST}$ (usually unobservable because its direct estimation requires special microscopic datasets like ours) can be inferred via the LMF predictions~\eqref{eq:quantitative_pred} and \eqref{eq:ACF_estimate_M_set}, where $\gamma$ and $c_0$ are observable even for public data. This is reminiscent of Einstein's theory for physical Brownian motions: Avogadro's number $N_A$ (unobservable) was indirectly estimated from the thermal fluctuations via the Einstein relation for the diffusion constant. The LMF theory can play a similar role in inferring microscopic financial parameters from financial fluctuations.

	The microscopic parameter set ($\alpha, N_{\ST})$ quantifies how the latent demand is hidden in the long term. For markets with small $\alpha$, the revealed liquidity on the limit-order book is insufficient for liquidity takers, and takers have no choice but to split their large metaorders into a longer series of child orders (see also \cite{BouchaudText} for a standard interpretation of the order-splitting behaviour from the viewpoint of practitioners). In this sense, markets with smaller $\alpha$ and large $N_{\ST}$ might not be liquid enough because many large institutional investors are waiting for the liquidity to replenish during their order-splitting. 
	This characteristic of liquidity has not been captured in practice through conventional metrics such as market spread (the difference between the best bid and ask prices), market depth (the typical volume size at the best prices), and market impact (the average price movement after a market order). Thus, the parameter set $(\alpha, N_{\ST})$ is a new measure quantifying how the market is potentially illiquid due to the hidden demand by large institutional investors. 

	Remarkably, successful strategy clustering was the key to our data analysis at the individual trader level in revealing the market ecology from a microscopic viewpoint. This research direction aligns with the previous literature~\cite{MantegnaEcology, SueshigeEcology1,SueshigeEcology2} proposing the need of market-ecology analyses. We believe that this direction of research holds promise, particularly for econophysics and sociophysics modelling~\cite{PhysRep2022} at it benefits from recent microscopic financial datasets.

\begin{acknowledgments}
	YS was supported by JST SPRING (Grant Number JPMJSP2110).
	KK was supported by JST PRESTO (Grant Number JPMJPR20M2), JSPS KAKENHI (Grant Numbers 21H01560 and 22H01141), and JSPS Core-to-Core Program (Grant Number JPJSCCA20200001).
	We greatly appreciate the data provision and careful review of this Letter by the JPX Group, Inc. We declare no financial conflict of interest. The JPX Group, Inc. provided the original data for this study without any financial support. We thank Richard Haase, PhD, from Edanz for editing a draft of this manuscript.
\end{acknowledgments}



\end{document}